\documentclass{PoS}
\usepackage{latexsym}
\usepackage{graphicx}
\usepackage{amsmath}
\usepackage{amssymb}

\usepackage{url}

\title{Hadronic vacuum polarization in finite volume using NNLO ChPT}

\ShortTitle{Hadronic vacuum polarization in finite volume using NNLO ChPT}

\author{\speaker{C. Aubin}\\Dept. of Physics \&
Engineering Physics, Fordham Univ., 
Bronx, NY 10458, USA\\
        E-mail: \email{caubin@fordham.edu}}

\author{T. Blum\\
        Dept. of Physics, Univ. of Connecticut, Storrs, CT 06269, USA}
\author{M. Golterman\\Dept. of Physics and Astronomy, San Francisco State Univ., San Francisco, CA 94132, USA
}
\author{C. Jung\\Physics Dept., Brookhaven National Laboratory, Upton, New York 11973, USA}
\author{S. Peris\\Dept. of Physics, Univ. Aut\`onoma de Barcelona, E-08193 Bellaterra, Barcelona, Spain}
\author{C. Tu\\ 
        Dept. of Physics, Univ. of Connecticut, Storrs, CT 06269, USA
}
\abstract{We present results for the leading hadronic 
contribution to 
the muon $g-2$ from configurations with 2+1+1 flavors 
of HISQ quarks. 
The ensembles have been generated by the MILC collaboration at 
three lattice spacings. Using the time-momentum representation 
of the electromagnetic current correlator, we calculate the 
finite volume effects up to next-to-next-to-leading-order in 
Chiral Perturbation Theory.}

\FullConference{37th International Symposium on Lattice Field Theory - Lattice2019\\
		16-22 June 2019\\
		Wuhan, China}

\begin{document}

\section{Introduction}\label{sec:intro}

One of the more promising quantities with which 
to test the Standard Model is the anomalous magnetic
moment of the muon $a_\mu = (g-2)/2$, given the
strong tension between the theoretical calculation (see,
for example, Ref.~\cite{Davier:2010nc})
and the experimental result \cite{Bennett:2006fi}. 
The experiment at Fermilab (E989), running now, is
expected to reduce the experimental uncertainty by 
a factor of four, and thus it is imperative for 
the theoretical calculation to obtain a similar
reduction in uncertainty. 

Given the improvements in 
lattice simulations of the leading 
hadronic contribution to the muon $g-2$ over the 
last decade or so, extracting this quantity with
the required precision from
a first-principles approach is seeming much more likely.
However, in addition to improving statistics on the
lattice data, many systematics 
must be understood and corrected for before a reliable
result can be obtained. 

We focus here primarily on the
systematics that enter when calculating the leading
hadronic contribution to the muon $g-2$ with staggered
quarks in a finite volume. 
To do so, we have calculated these effects to next-to-next-to
leading order (NNLO) 
in chiral perturbation theory 
(ChPT). A more detailed description of our results
can be found in Ref.~\cite{Aubin:2019usy}.

\section{Simulation details}\label{sec:details}

The leading hadronic contribution to the muon
$g-2$ can be obtained from the expression
\begin{equation}\label{eq:amu}
	a_\mu^{\rm HVP} 
	= 
	4 \alpha^2
	\int_0^\infty dq^2 f(q^2) \hat\Pi(q^2),
\end{equation}
where $f(q^2)$ is defined in Ref.~\cite{Blum:2002ii},
and $\hat \Pi(q^2) = \Pi(q^2) - \Pi(0)$ is the
subtracted hadronic vacuum polarization, 
coming from the Fourier transform of the vector
two-point function (we use the conserved vector
current here). 
In this work, we
use the time-momentum representation:
\begin{eqnarray}\label{eq:corr}
	\Pi(q^2) -\Pi(0) = \sum_t \left(
	\frac{\cos{qt} -1}{q^2} +\frac{1}{2} t^2
	\right) C(t),\quad 
%\label{eq:corr}
	C(t) = \frac{1}{3}\sum_{\vec x,i}\langle 
	j^i(\vec x, t)j^i(0)\rangle,
 \end{eqnarray}
 where $C(t)$ is the Euclidean time correlation function, 
 averaged over spatial directions. Eq.~(\ref{eq:amu}) becomes
$a_{\mu}^{\rm HVP}(T) = \sum_{t} w(t) C(t)$,
with the weight
\begin{equation}\label{eq:kernel}
 	w(t) = 4 \alpha^2 \int_{0}^{\infty} {d \omega^2}
 	f(\omega^2)\left[\frac{\cos{\omega t} -1}{\omega^2} 
 	+  \frac{t^2}{2}\right].
\end{equation}
The weight is sometimes modified by replacing 
the continuum Euclidean momentum-squared with 
its lattice version $\hat w(t)$, where the $\omega^2$ in the 
denominator of the first term in square brackets 
is replaced with $[2\sin(\omega/2)]^2$ \cite{Blum:2018mom}. 

In order to calculate the correlator in Eq.~(\ref{eq:corr}), 
we implement the noise reduction techniques developed by
RBC/UKQCD \cite{Blum:2012uh,Blum:2018mom} including a 
combination of all-mode and full volume low-mode averaging. 
The precise details of the implementation of these 
techniques to the staggered Dirac operator are discussed
in Ref.~\cite{Aubin:2019usy}. 

The ensembles used are 2+1+1 HISQ configurations generated by the MILC 
collaboration \cite{Bazavov:2014wgs}, 
and are listed in Ref.~\cite{Aubin:2019usy}. We simulated at three lattice 
spacings ($a\approx 0.06,0.09,$ and 0.12 fm) at 
approximately physical pion masses. % in MeV, 
The volumes all have a 
spatial volume of around $(5.5~\rm{fm})^3$, and 
$m_\pi L\approx 3.7 - 3.9$. 
For the two
coarser ensembles we 
generated 3000 eigenvectors for the low-modes, 
however only 2000 were used on the
finest ensemble due to computational limitations.

\section{Finite-volume chiral perturbation theory}\label{sec:chpt}

In order to study the leading finite-volume effects we calculate
the vector correlator to two loops 
(NNLO) in ChPT. There are several strategies we could use for this. 
One option is to first extrapolate our
 results to the continuum and then
correct for the finite-volume using continuum ChPT. The
other option would be to first use staggered 
ChPT \cite{Aubin:2003mg} in a finite volume and then 
extrapolate to the continuum. Given that our pion 
masses and our physical volumes are not exactly equal, 
the second approach would be better able to take these
differences into account. However, this would require
applying staggered ChPT to two loops.

Instead we choose a hybrid approach. First we calculate
the corrections using staggered ChPT at one loop for
each ensemble, and then extrapolate to the continuum. 
At this point, we calculate the NNLO \emph{continuum} 
finite volume corrections, without the use of 
(or need for) the staggered taste-breaking corrections.
There will still be a small systematic effect coming
from the slight mistunings of the pion masses and
volumes, however it will be much smaller than 
if we were to extrapolate to the continuum first, 
and then apply the complete NLO+NNLO continuum 
ChPT to correct for finite-volume effects.

In Euclidean space we have performed a relatively
straightforward calculation in the 
time-momentum representation to obtain $C(t)$ to NNLO:
\begin{eqnarray}\label{CtNNLOdimreg}
	C(t) & = & 
	\frac{10}{9}\frac{1}{3}\Biggl\{\frac{1}{L^d}
	\sum_{{\vec p}}\frac{{\vec p}^2}{E_p^2}\,e^{-2E_p t}
	\Biggl[1-\frac{2}{F^2}\frac{1}{L^d}\sum_{\vec k}\left(\frac{1}{2E_k}\right) - 
	\frac{8({\vec p}^2+m_\pi^2)}{F^2}\,\ell_6\Biggr]
	\\&&
	\phantom{\frac{10}{9}\frac{1}{3}\,e^2\Biggl(}
	 + \frac{1}{2dF^2}\frac{1}{L^{2d}}
	 \sum_{{\vec p},{\vec k}}
	\frac{{\vec p}^2{\vec k}^2}{E_p^2E^2_k}
	\frac{E_k e^{-2E_p t}-E_p e^{-2E_k t}}{{\vec k}^2-
	{\vec p}^2}
	\Biggr\}\ ,
	\nonumber
\end{eqnarray}
where we have defined $E_p  = \sqrt{m_\pi^2+{\vec p}^2}$.
The sums over $\vec p$ and $\vec k$ are over 
the momenta $2\pi{\vec n}/L$ with ${\vec n}$ a three-vector
of integers in a box with periodic boundary conditions. 
We define the renormalized $\ell_6^r$ by
\begin{equation}\label{l6ren}
	\ell_6 = \ell_6^r(\mu)
	-\frac{1}{3}\,\frac{1}{16\pi^2}
	\left(\frac{1}{\epsilon}-\log\mu - 
	\frac{1}{2}(\log{(4\pi)}-\gamma+1)\right)\ ,
\end{equation}
to take the limit $d=3+\epsilon\to 3$ in Eq.~(\ref{CtNNLOdimreg}) to
obtain a finite result for $C(t)$. 

Looking at the expression in Eq.~(\ref{CtNNLOdimreg}) term-by-term,
we can obtain expressions for the NLO and NNLO finite volume
corrections~\cite{Aubin:2019usy}, defined by
\begin{equation}
\Delta a_\mu^{\rm HVP}
	=
	\left[\lim_{L\to\infty} a_\mu^{\rm HVP}(L)\right]
	 -
 	a_\mu^{\rm HVP}(L)\ .
\end{equation}
Using the parameter values from Ref.~\cite{Aubin:2019usy}, we obtain
the results for $\Delta a_\mu^{\rm HVP}$ at NLO (coming from the ``1''
in square brackets 
on the first line of Eq.~(\ref{CtNNLOdimreg})), shown in the second
column in Table~\ref{tab:deltaamu}.
From the same expression used for the NLO corrections, we can 
use the staggered pion spectrum to include the effects
of the different taste masses in finite volume
(third column) and we can
calculate the 
effect of taste breaking in the pion masses 
in the infinite volume limit to NLO in ChPT, shown in the fourth
column of Table~\ref{tab:deltaamu}.

\begin{table}[htp]
\begin{center}
\begin{tabular}{|c|c|c|c|c|c|}
\hline
 $a$ (fm) & NLO  & taste (lattice)& taste (cont) & NNLO & total \\
\hline
 0.12121(64) & 18.08 & 2.1 & 51.6 & 7.40 & $25.5\pm3.0$ \\
 0.08787(46) & 21.60 & 6.9 & 34.2 & 9.01 & $30.6\pm3.8$ \\
 0.05684(30) & 20.59 & 15.6 &  9.5 & 9.13 & $29.7\pm4.0$ \\
\hline
\end{tabular}
\end{center}
\caption{ChPT FV corrections to the muon $a_\mu$,
$\Delta a_\mu^{\rm HVP}$, 
in units of $10^{-10}$. The columns are 
discussed in the text.}
\label{tab:deltaamu}
\end{table}%

The NNLO continuum finite-volume corrections
come from an application of the Poisson summation 
formula to the remaining terms in Eq.~(\ref{CtNNLOdimreg}),
and the results 
are listed in the fifth column of Table~\ref{tab:deltaamu}. 
Finally, we obtain the total NLO+NNLO corrections and 
show these in the final column of Table~\ref{tab:deltaamu}. 
The errors are determined by assuming the omitted corrections are smaller
than the NNLO terms by the same factor ($\sim0.4$--$0.45$) as the NNLO
corrections are compared to the NLO contributions.
Finally, the effects of taste-splittings in the staggered pion spectrum
have been taken into account, and the details 
can be seen in Ref.~\cite{Aubin:2019usy}.

\section{Results \& Conclusions}\label{sec:conc}

\begin{figure}[htbp]
\begin{center}
\includegraphics[width=0.45\textwidth]{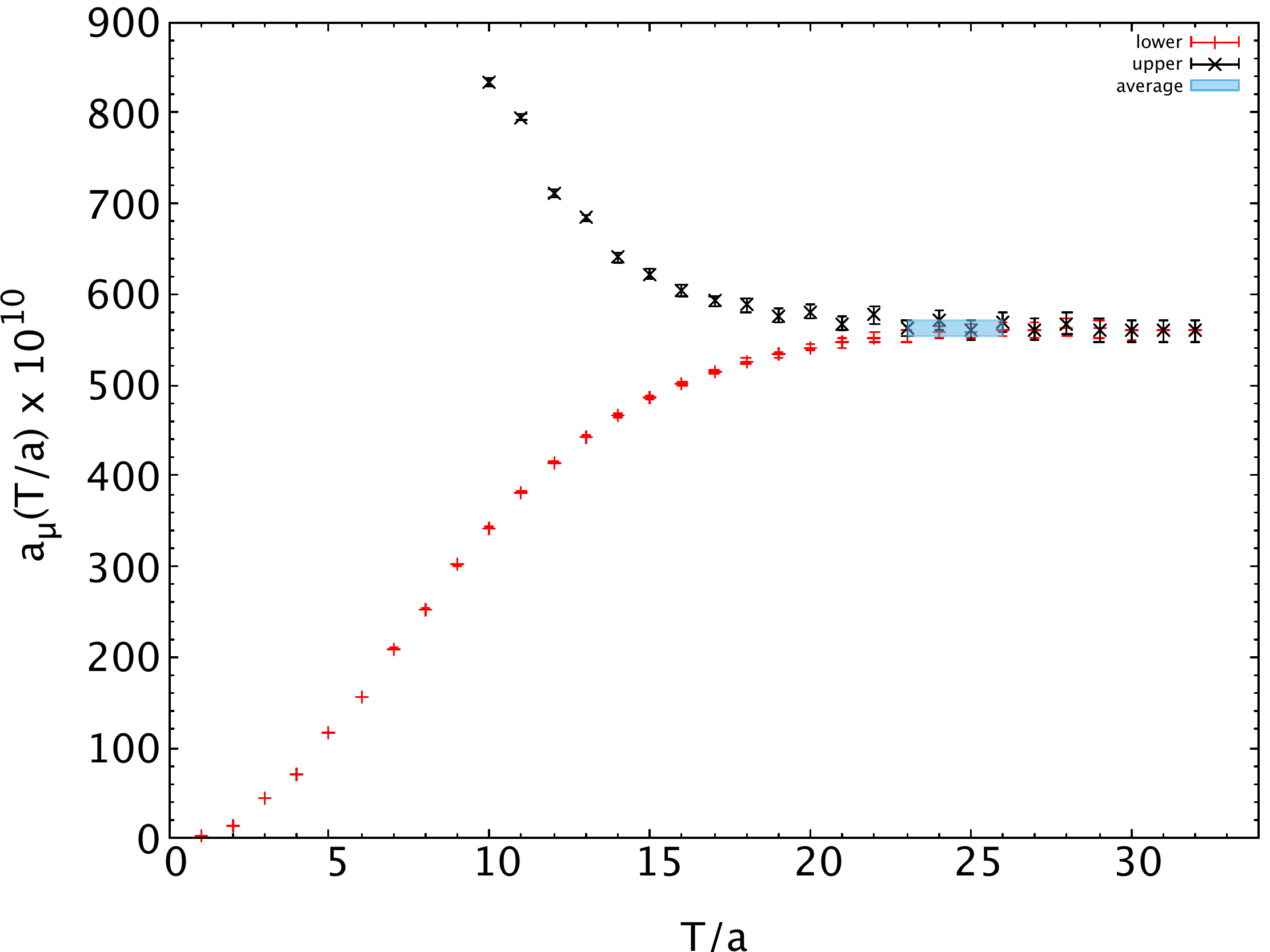}
\includegraphics[width=0.45\textwidth]{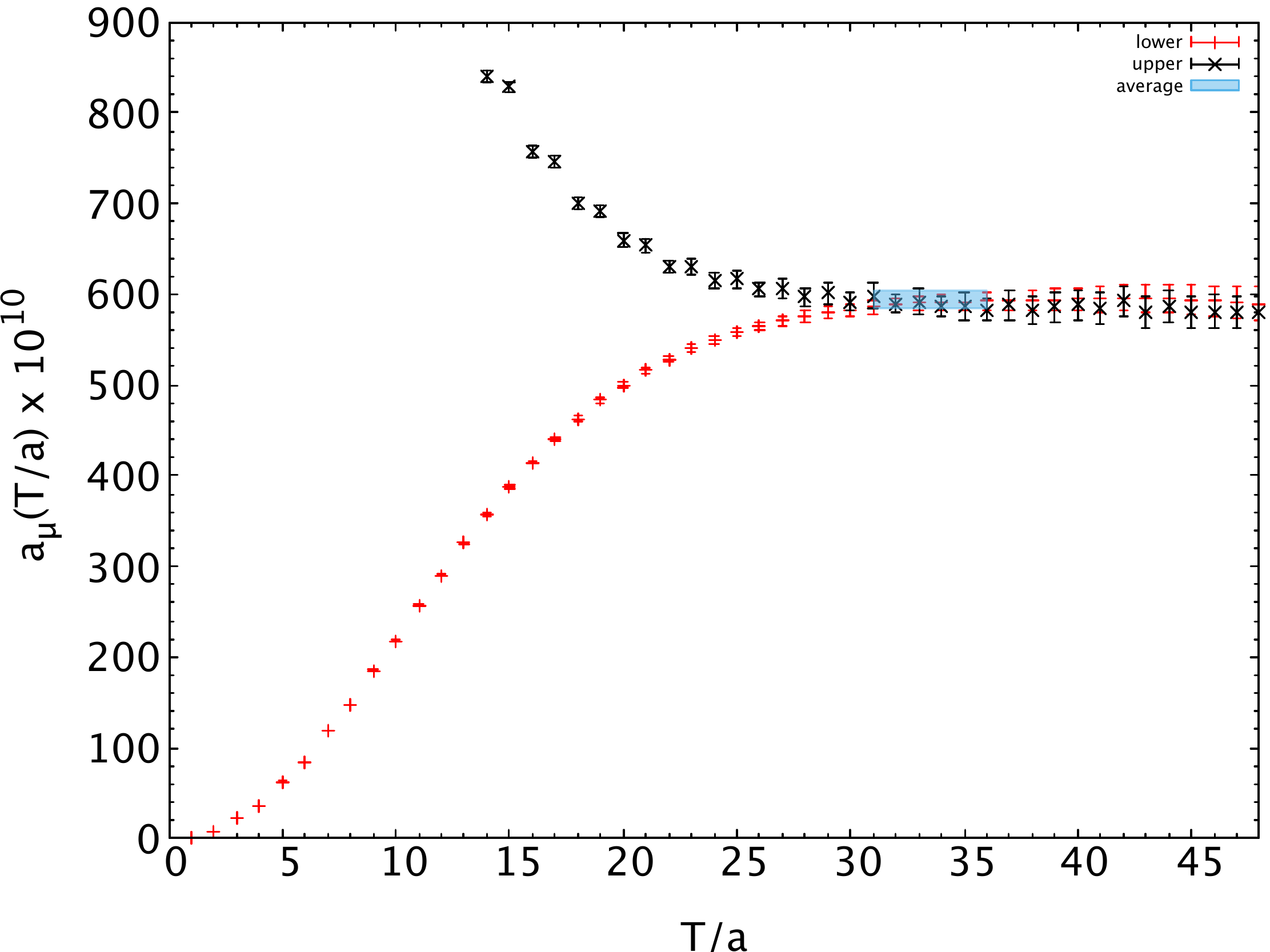}
\includegraphics[width=0.45\textwidth]{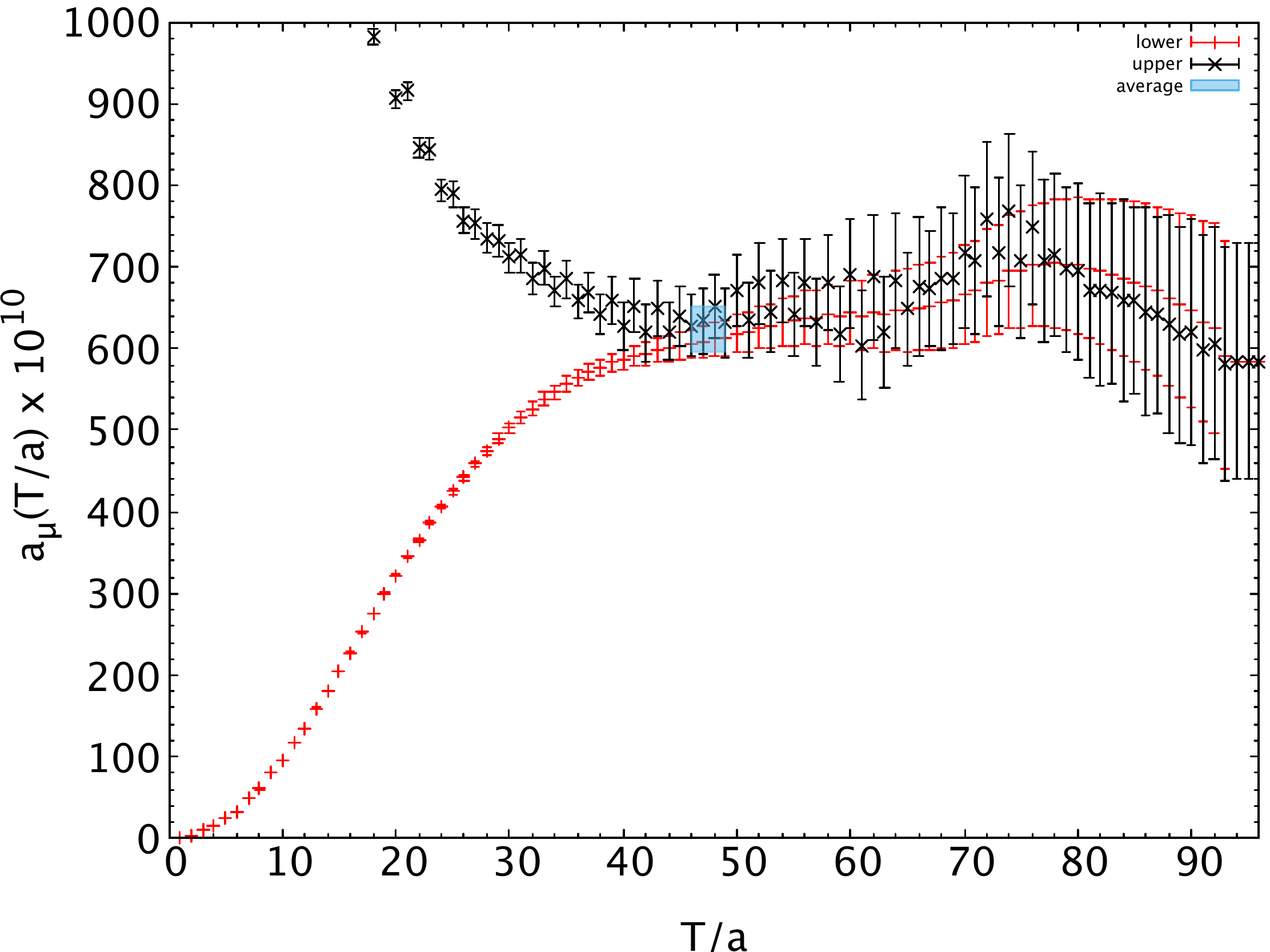}
\caption{Bounding method for total contribution to 
the muon anomaly, using the weighting function $w$. Clockwise from the top we have the
$48^3$, $64^3$, and $96^3$ ensembles.}
\label{fig:bound}
\end{center}
\end{figure}

In addition to using the techniques discussed above to reduce
statistical noise, we additionally apply the bounding method of 
Refs.~\cite{Blum:2018mom,Borsanyi:2017zdw} where we set upper and lower 
bounds on the
correlator for $t>T$: $C(t)=0$ and $C(t)=C(T) e^{-E_0(t-T)}$ 
respectively, where the lowest energy state in the vector channel is 
$E_0=2\sqrt{m_\pi^2+(2\pi/L)^2}$. 
At sufficiently large $T$ the bounds overlap, and an estimate 
for $a_\mu$ can be made which may be more precise than simply summing 
over the noisy long-distance tail. 

In Fig.~\ref{fig:bound} results for the bounding
method are shown for each ensemble. We calculate central values 
for $a_\mu$ by averaging over a suitable range where $T$ is 
large enough for the bounds to overlap but not so large 
that statistical errors blow up. The ranges used were 
2.7-3.2 fm for the $48^3$ and $64^3$ ensembles, and 2.6-2.8 fm for $96^3$ 
ensemble, with statistical errors computed using the jackknife method.

\begin{table}[htp]
\begin{center}
\begin{tabular}{|c|c|c|c|c|}
\hline
 $a$ (fm) & lattice value & FV corr. & FV + taste corr. & FV+taste+$m_\pi$ corr. \\
\hline
 0.12121(64) &562.1(8.4) & 564.2(8.4)  &615.8(8.4)& 613.6(8.4)\\
 0.08787(46) &594.8(10.4)& 601.7(10.4)  &  635.9(10.4)& 630.2(10.4)\\
 0.05684(30) &623.1(27.5)& 638.7(27.5) & 648.2(27.5) & 647.1(27.5)\\
 \hline
0 &  & 648.3(20.0) & 657.9(20.0)& 651.1(20.1)\\
\hline
\end{tabular}
\end{center}
\caption{HVP contributions to the muon $a_\mu$, 
in units of $10^{-10}$, including ChPT corrections. The columns are 
discussed in the text.}
\label{tab:resultscorrected}
\end{table}%

\begin{figure}[htbp]
\begin{center}
\includegraphics[width=0.45\textwidth]{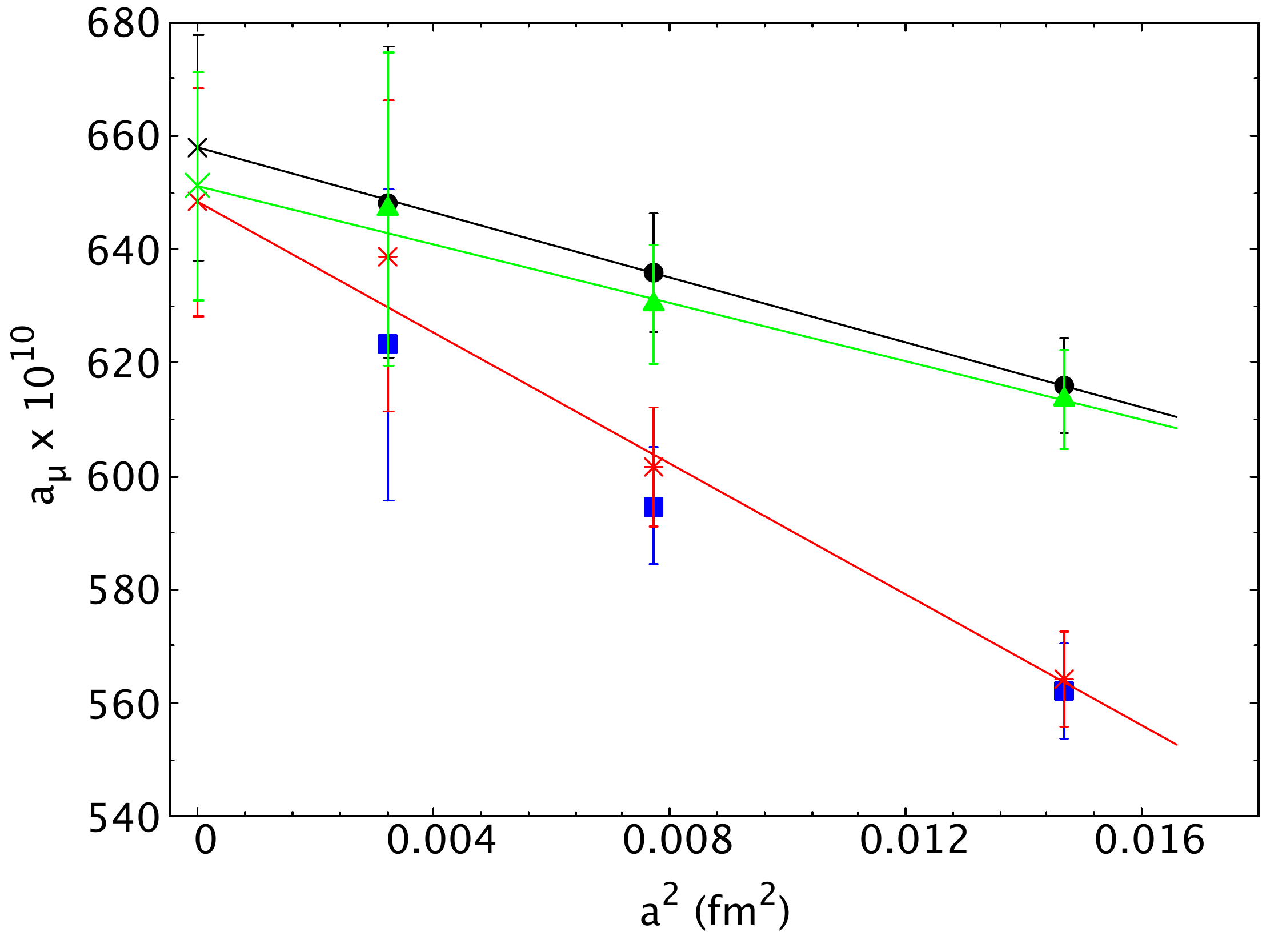}
\includegraphics[width=0.45\textwidth]{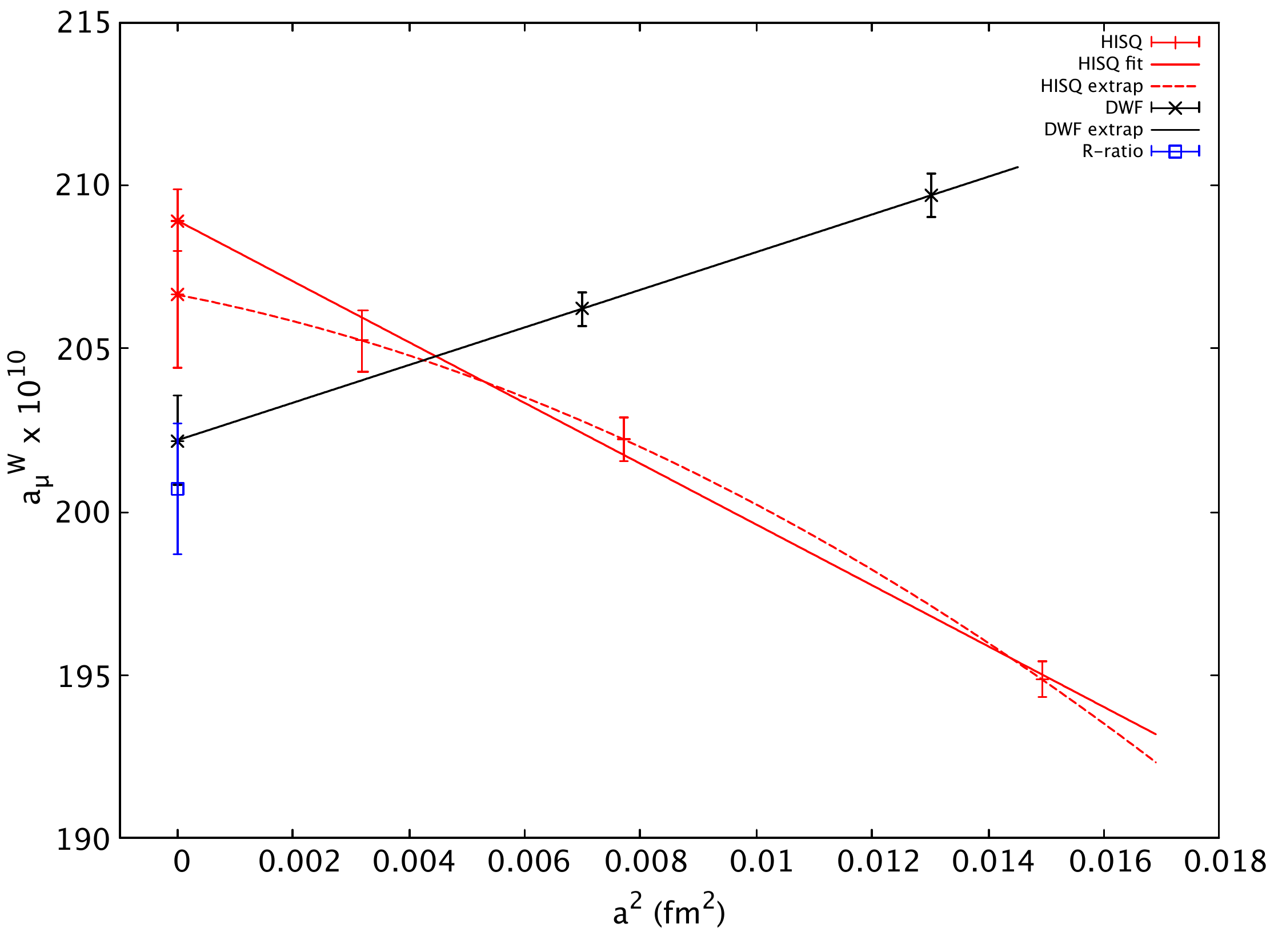}
\caption{Left: Continuum limit after correcting the data 
according to columns 3, 4, and 5 (bursts, circles, and 
triangles, respectively) of Table~\ref{tab:resultscorrected} with
the uncorrected data (squares) shown for comparison.
Right: Continuum limit combined with the window method as described in the text. }
\label{fig:conttab}
\end{center}
\end{figure}

The corrected results are tabulated in Table~\ref{tab:resultscorrected}
and shown with the continuum limits in Fig.~\ref{fig:conttab} (left figure). 
In Table~\ref{tab:resultscorrected}, the 
second column includes the results from the bounding method, 
the third column includes the finite-volume corrections 
of Ref.~\cite{Aubin:2019usy}, while the fourth column also includes 
the infinite-volume taste corrections in Table~\ref{tab:deltaamu}.
The fifth column adjusts the values shown in the fourth 
column to a common pion mass of 135~MeV using NLO ChPT, to account for
the small mistunings of the pion mass.
Continuum extrapolated values of each column are shown in the 
last row. The left plot of Fig.~\ref{fig:conttab} shows the continuum
limits taken to get the results shown in Table~\ref{tab:resultscorrected}.

After taking all of the corrections discussed into account, we obtain
for our final result
\begin{equation}\label{eq:finalResult}
	a_\mu^{\rm HVP} 
	=
	(659\pm 20 \pm 5\pm 5 \pm 4)\times 10^{-10}
	= 
	659(22)\times 10^{-10}
\end{equation}
where the errors quoted are, respectively, statistical, continuum 
extrapolation, scale setting, and higher orders in ChPT, and the final
result shows them added in quadrature. 

\begin{figure}[htbp]
\begin{center}
\includegraphics[width=0.6\textwidth]{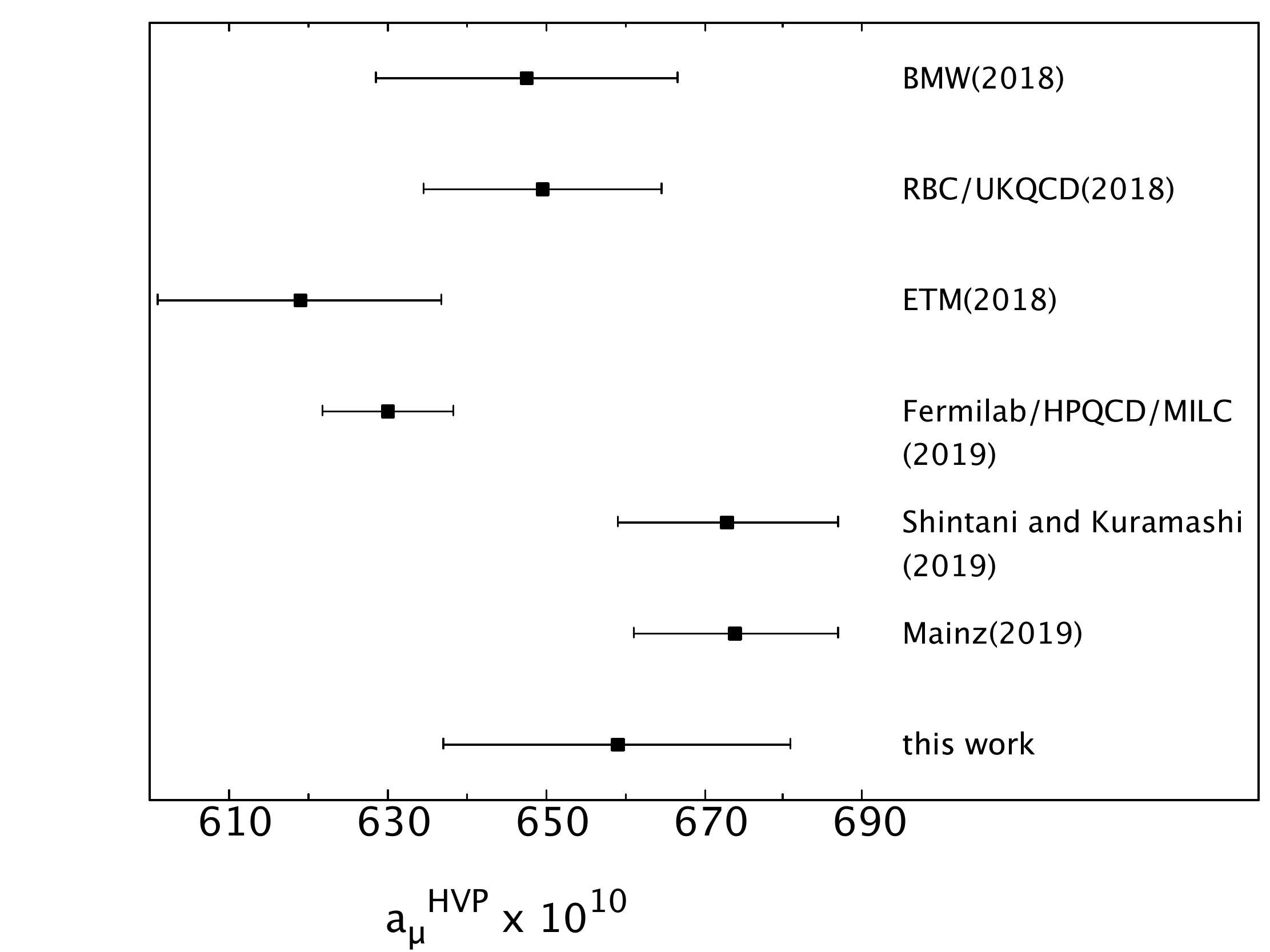}
\caption{Contributions to the $g-2$ from the connected 
light quark vacuum polarization from recent 
publications \cite{Borsanyi:2017zdw} (BMW), \cite{Blum:2018mom} (RBC/UKQCD), \cite{Giusti:2018mdh} (ETM), \cite{Davies:2019efs} (Fermilab/HPQCD/MILC), \cite{Shintani:2019wai} (Shintani and Kuramashi), \cite{Gerardin:2019rua} (Mainz).}
\label{fig:amu lit}
\end{center}
\end{figure}

To explore a more precise comparison with other results, 
we adopt the window method of Ref.~\cite{Blum:2018mom}: 
$%\begin{eqnarray}
a_\mu^{W} = 2\sum_{t=0}^{T/2} C(t) w(t) (\Theta(t,t_0,\Delta)-\Theta(t,t_1,\Delta))$, 
with $\Theta(t,t',\Delta) = \frac{1}{2} (1+\tanh((t-t')/\Delta))$,
where $t_1-t_0$ is the size of the window and $\Delta$ is a 
suitably chosen width that smears out the window at either edge. We 
choose windows to avoid both lattice artifacts at short distance 
and large statistical errors at long distance. Results for several 
windows and both weighting functions are tabulated in Ref.~\cite{Aubin:2019usy}.

In Fig.~\ref{fig:conttab} (right figure), we show an example continuum 
limit combined with
the window method with $t_0=0.4$ fm, $t_1=1$ fm,
$\Delta=0.15$.\footnote{We note that since this talk was given,
we have updated this figure to better compare with the domain wall
fermion calculation, and this will appear in the latest version
of Ref.~\cite{Aubin:2019usy}.}
Squares (crosses) correspond to uncorrected data points with 
weighting function $w$ ($\hat w$); filled circles are taste-breaking 
corrected to NLO of $w$ data points.  Solid curves show linear 
fits in $a^2$; all three agree very well in the continuum limit. 
Dashed curves denote a fully constrained parametrization 
(no degrees of freedom) using both $a^2$ and $a^4$ terms. Additionally 
we include the recent RBC/UKQCD computation using domain wall fermions, as
the results should agree in the continuum limit up to small systematics.

We also show the corresponding dispersive/$e^+e^-$ value, using the 
R-ratio compilation of Ref.~\cite{Keshavarzi:2018mgv}. The 
largest difference is about $7\times 10^{-10}$, or roughly 1~\% 
of the total HVP contribution to $a_\mu$.
Given the uncertainties it is difficult to conclude there is a 
significant discrepancy, though the spread seems uncomfortably large. 
A third, smaller, lattice spacing ensemble is being generated by the 
RBC/UKQCD collaborations~\cite{rbcukqcd}, which could firmly establish whether 
or not a discrepancy exists. The window method is a useful approach to 
cross-check different calculations using the most precise data available 
for each.

\bibliographystyle{JHEP}   % if natbib is available

\bibliography{refs2}

\end{document}